\documentclass[a4paper,UKenglish, thm-restate]{lipics-v2021}

\title{Combinatorial and Algorithmic Aspects of Monadic Stability} 

\usepackage[all,defaultlines=3]{nowidow}

\titlerunning{Combinatorial and Algorithmic Aspects of Monadic Stability} 

\author{Jan Dreier}{TU Wien, Austria}{dreier@ac.tuwien.ac.at}{https://orcid.org/0000-0002-2662-5303}{}

\author{Nikolas M\"ahlmann}{University of Bremen, Germany}{maehlmann@uni-bremen.de}{https://orcid.org/0000-0003-3657-7736}{}

\author{Amer E.~Mouawad}{American University of Beirut, Lebanon and\\ University of Bremen, Germany}{aa368@aub.edu.lb}{https://orcid.org/0000-0003-2481-4968}{}

\author{Sebastian Siebertz}{University of Bremen, Germany}{siebertz@uni-bremen.de}{https://orcid.org/0000-0002-6347-1198}{}{}

\author{Alexandre Vigny}{University of Bremen, Germany}{vigny@uni-bremen.de}{https://orcid.org/0000-0002-4298-8876}{}{}

\authorrunning{J. Dreier, N. M\"ahlmann, A. E. Mouawad, S. Siebertz, A. Vigny} 

\Copyright{Jan Dreier, Nikolas M\"ahlmann, Amer E. Mouawad, Sebastian Siebertz, and Alexandre Vigny} 

\ccsdesc[500]{Theory of computation~Logic}
\ccsdesc[500]{Theory of computation~Finite Model Theory}
\ccsdesc[500]{Mathematics of computing~Combinatorics}

\keywords{Stability/classification theory, structural graph theory, first-order logic, indiscernible sequences} 

\category{} 

\relatedversion{} 

\funding{This paper is a part of a project that has received funding from the German
	Research Foundation (DFG) with grant agreement
	No 444419611.}

\nolinenumbers 

\hideLIPIcs  

\EventEditors{Florin Manea and Alex Simpson}
\EventNoEds{2}
\EventLongTitle{30th EACSL Annual Conference on Computer Science Logic (CSL 2022)}
\EventShortTitle{MFCS 2022}
\EventAcronym{CSL}
\EventYear{2022}
\EventDate{February 14--19, 2022}
\EventLocation{G\"{o}ttingen, Germany (Virtual Conference)}
\EventLogo{}
\SeriesVolume{216}
\ArticleNo{20}
\usepackage[noabbrev,capitalise,nameinlink]{cleveref}

\crefname{observation}{Observation}{Observation}
\crefalias{enumi}{observation}
\crefname{claim}{Claim}{Claim}

\usepackage{hyperref}
\usepackage{enumitem}
\usepackage{mathrsfs}

\numberwithin{table}{section}
\numberwithin{lemma}{section}
\numberwithin{theorem}{section}
\numberwithin{claim}{section}
\numberwithin{corollary}{section}
\numberwithin{definition}{section}

\usepackage{amsmath}
\usepackage{amssymb}
\usepackage{amsthm}
\usepackage{thm-restate}
\usepackage{mathtools}

\usepackage{tcolorbox}

\usepackage{float}
\usepackage{wrapfig}

\usepackage{tikz}
\usepackage{cancel}
\usepackage{mdframed}
\usepackage{xspace}
\usepackage[textsize=tiny]{todonotes}

\newcommand{\Cc}[0]{\ensuremath{\mathscr{C}}\xspace}
\newcommand{\Dd}[0]{\ensuremath{\mathscr{D}}\xspace}

\newcommand{\CC}[0]{\mathrm{\mathscr{C}}}
\newcommand{\DD}[0]{\mathrm{\mathscr{D}}}
\newcommand{\NN}[0]{\mathrm{\mathbb{N}}}

\newcommand{\GG}{\mathcal{G}}
\newcommand{\KK}{\mathcal{K}}

\renewcommand{\phi}{\varphi}

\newcommand{\Oof}{\mathcal{O}}

\DeclareMathOperator{\dens}{\rm dens}

\newcommand{\forkindep}[1]{%
  \mathrel{
    \mathop{
      \vcenter{
        \hbox{\oalign{\noalign{\kern-.3ex}\hfil$\vert$\hfil\cr
              \noalign{\kern-.7ex}
              $\smile$\cr\noalign{\kern-.3ex}}}
      }
    }\displaylimits_{#1}
  }
}

\newcommand{\sdc}[2]{\mathrm{sd}_{#1}(K_{#2})}

\begin{document}

\maketitle

\begin{abstract}
Nowhere dense classes of graphs are classes of sparse graphs with rich structural and algorithmic properties, however, they fail to capture even simple classes of dense graphs. 
Monadically stable classes, originating from model theory, 
generalize nowhere dense classes and close them under transductions, i.e.\ transformations defined by colorings and simple first-order interpretations.
In this work we aim to extend some combinatorial and algorithmic properties of nowhere dense classes to monadically stable classes of finite graphs. We prove the following results. 
\vspace{-2mm}

\begin{itemize}
\item In monadically stable
classes the Ramsey numbers $R(s,t)$ are bounded from above by
$\mathcal{O}(t^{s-1-\delta})$ for some $\delta>0$, improving the
bound $R(s,t)\in \mathcal{O}(t^{s-1}/(\log t)^{s-1})$ known for general graphs and the bounds known for $k$-stable graphs when $s\leq k$. 
\item For every monadically stable class $\CC$ and every integer~$r$, there exists $\delta > 0$ such that every graph~$G \in \CC$ that contains an~$r$-subdivision of the biclique $K_{t,t}$ as a subgraph also contains~$K_{t^\delta,t^\delta}$ as a subgraph. This generalizes earlier results for nowhere dense graph classes.
\item We obtain a stronger regularity lemma for monadically stable classes of graphs.
\item Finally, we show that we can compute polynomial kernels for the independent set and dominating set problems in powers of nowhere dense classes. Formerly, only fixed-parameter tractable algorithms were known for these problems on powers of nowhere dense classes.
\end{itemize}
\end{abstract}

\section{Introduction}

The notion of nowhere density was introduced by Ne\v{s}et\v{r}il and Ossona de Mendez~\cite{nevsetvril2011nowhere} and provides a very robust notion of uniform sparsity of graphs. These classes can be characterized in various ways, leading to a rich combinatorial theory and enabling the design of efficient algorithms for problems that are hard in general. For example, the (distance-$r$) dominating set problem~\cite{DawarK09} and more generally, the first-order model-checking problem~\cite{grohe2017deciding}, is fixed-parameter tractable on nowhere dense graph classes.

Unfortunately, the techniques developed for nowhere dense classes are not applicable on dense graphs. However, there has been recent success of defining notions of \emph{structural sparsity} by considering first-order transductions of sparse classes. Intuitively, a first-order transduction transforms graphs by first non-deterministically adding colors and then reinterpreting the edge relation using a first-order formula~$\phi(x,y)$, see e.g.~\cite{dreier2022treelike,gajarsky2020new,gajarsky2020first,nevsetvril2016structural,
nevsetvril2022structural}. 
This has led to the notion of \emph{structurally nowhere dense} classes, which are first-order transductions of nowhere dense classes. If $\Cc$ is a nowhere dense class, then for example the class of all complements of graphs from $\Cc$ or for an integer $d$ the class of all $d$th powers of graphs from $\Cc$ forms a structurally nowhere dense class. 

It has been conjectured that structurally nowhere these classes coincide with \emph{monadically stable} classes (see e.g.~\cite{nevsetvril2021rankwidth}), which have long been studied in model theory~\cite{baldwin1985second}. This conjecture is based on the results of Adler and Adler~\cite{adler2014interpreting}, showing that every nowhere dense class is monadically stable and in fact for monotone classes (i.e.\ classes closed under taking subgraphs) the notions of nowhere density and monadic stability coincide. However, monadically stable classes are vastly more general than nowhere dense classes. Further indications for the truth of the conjecture is given by a recent characterization of monadically stable classes in terms of \emph{flip-wideness}~\cite{dreier2022iaw}, which generalizes the notion of uniform quasi-wideness, which characterizes nowhere dense classes. 

So far, the main tool for the study monadically stable classes has been \emph{indiscernible sequences}~\cite{dreier2022iaw}. In our restricted context we focus on indiscernible sequences of vertices, that is, sequences $(v_i)_{1\leq i\leq n}$ of vertices such that all (increasing tuples of) vertices from the sequence satisfy the same first-order formulas. For example, an indiscernible sequence of vertices in a graph will induce an independent set or a clique. Furthermore, indiscernible sequences impose strong structural properties on their neighborhoods, which we exploit in our combinatorial and algorithmic study of monadically stable classes. 

In the remainder of this section we describe our contribution in detail. 

\smallskip
\noindent
\textbf{Ramsey numbers in monadically stable classes of
  graphs.}  Ramsey's theorem is a foundational result in
combinatorics, stating in one of its graph theoretic formulations that
in every sufficiently large graph one either finds a large clique or a
large independent set. The Ramsey number $R(s,t)$ denotes the smallest
number of vertices that a graph must have such that one can either
find a clique of size $s$ or an independent set of size $t$.  Focusing
on the case that $s$ is fixed and $t$ is growing, folklore bounds give
us $R(s,t) \le { s+t-2 \choose s-1} \in \mathcal{O}_s(t^{s-1})$ (the
subscript $s$ in the asymptotic notation indicates that constant
factors depending on $s$ are hidden).  For $s=1,2$ this bound is
trivial and tight.  The classical result of Ajtai, Koml{\'o}s and
Szemer{\'e}di~\cite{ajtai1980note} improves this for $s \ge 3$ to
$R(s,t) \in \mathcal{O}_s(t^{s-1})/(\log t)^{s-2}$.  We can rephrase
this by saying that every $K_s$-free $n$-vertex graph with $s \ge 3$
contains an independent set of order
$\Omega_s\left(n^{1/(s-1)}\cdot(\log n)^{(s-2)/(s-1)}\right)$.  For
every stable class $\Cc$ there exists an integer~$k$ (the ladder
index, see~\cref{sec:prelims} for formal definitions) such that every
$K_s$-free $n$-vertex graph $G\in \Cc$ with $s\geq k-1$ contains an
independent set of order $\Omega((n/k)^{1/(k+1)})$, strongly improving
the general bound by removing the dependence on $s$ from the
exponent. These bounds are implicit in~\cite{regularity_lemmas} and
made explicit in~\cite[Observation 6.12]{oyestable}.  However, on
stable graph classes for $s<k-1$ the classical bounds of Ajtai et al.\
are the currently best known.
%
Things look better on structurally nowhere dense classes:
For every such class~$\Cc$, $\epsilon > 0$ and integer $s$, it holds
that every $K_s$-free $n$-vertex graph from $\Cc$ contains an
independent set of size $\Omega_{s,\epsilon,\Cc}(n^{1-\epsilon})$.
This follows from the fact that graphs from structurally
nowhere dense classes can be partitioned into
$\mathcal{O}_{\Cc,\epsilon,s}(n^\epsilon)$ parts, each of which
induces a cograph~\cite{dreier2022treelike}, and since cographs are
perfect, they satisfy $R(s,t)=st$.  (To be more precise, it is proved
in~\cite{dreier2022treelike} that each of the parts has bounded
shrubdepth and it is proved in~\cite{nevsetvril2020linear} that every
graph of bounded shrubdepth can be decomposed into a bounded number of
cographs.)

In light of the conjecture that every monadically stable class is
a transduction of a nowhere dense class, we conjecture that
also for every monadically stable class $\Cc$, $\epsilon > 0$ and
integer $s$, it holds that every $K_s$-free $n$-vertex graph from
$\Cc$ contains an independent set of size
$\Omega_{\Cc,\epsilon,s}(n^{1-\epsilon})$.
We were unable to prove this conjecture, however, we were able to
improve the bounds for monadically stable classes with ladder index
$k$ and $s<k-1$ by proving the following. If $\Cc$ is a $K_s$-free
monadically stable class for some fixed $s \ge 3$, then there exists a
real $\delta>0$ such that in every $n$-vertex graph from $\Cc$ we can
find an independent set of size $\Omega_{\Cc}(n^{1/(s-1)+\delta})$.
This implies that in $\Cc$ we have
$R(s,t)\in \mathcal{O}_{\Cc}(t^{s-1-\delta})$ (\Cref{remark:Rst}
below).  To put this into perspective, Ajtai et al.'s classical result
improves the size of an independent in $K_s$-free graphs by a factor
$\Theta((\log n)^{(s-2)/(s-1)})$ compared to the folklore bounds of
$\Omega(n^{1/(s-1)})$ and our result improves it by a factor
$\Theta(n^\delta)$ for monadically stable graph classes.

\smallskip
\noindent
\textbf{Large bicliques in monadically stable classes of
  graphs.}  An $r$-subdivision of a graph~$H$ is a graph obtained by
replacing the edges of $H$ by paths of length (exactly) $r+1$
(containing $r$ inner vertices). A class $\Cc$ of graphs is nowhere
dense if and only if there exists a function~$f$ such that for every
$r\geq 0$ the $r$-subdivision of a biclique (complete bipartite graph)
$K_{f(r),f(r)}$ with parts of size $f(r)$ is excluded as a
subgraph~\cite{nevsetvril2012sparsity}.  Hence, every class that is
not nowhere dense must contain for some value of $r$ arbitrarily large
$r$-subdivided bicliques.  It follows from~\cite{dvovrak2018induced}
that every monadically stable class that is not nowhere dense even
contains arbitrarily large bicliques as subgraphs.  We demonstrate
that these two quantities (size of $r$-subdivided bicliques and size
of bicliques) are polynomially bounded in monadically stable classes,
that is, we prove that for every monadically stable class $\CC$ and
every $r>0$, there exist an integer $m$ and a real $\delta > 0$, such
that for every $t\geq m$ it holds that every graph $G \in \CC$ that
contains an $r$-subdivision of~$K_{t,t}$ as a subgraph, also contains
$K_{t^\delta,t^\delta}$ as a subgraph.

\smallskip
\noindent
\textbf{Regularity lemma for monadically stable classes of graphs.}
Szemer\'edi's regularity lem\-ma~\cite{sz} is a celebrated result in
extremal graph theory.  Let us give the required formal definitions to
formulate the lemma.  Let $G$ be a graph and let $A,B\subseteq V(G)$
be two disjoint non-empty subsets of vertices. We write $E(A,B)$ for
the set of edges with one end in $A$ and the other end in $B$. We
define the \emph{density} of the pair $(A,B)$ as
$\dens(A,B)\coloneqq {|E(A,B)|}/{|A| |B|}$.  Let $\epsilon>0$, let $G$
be a graph and let $A,B\subseteq V(G)$ be two disjoint non-empty
subsets of vertices. We call the pair $(A,B)$
\emph{$\epsilon$-regular} if, for all subsets $A'\subseteq A$ and
$B'\subseteq B$ with $|A'|\geq \epsilon|A|$ and
$|B'|\geq \epsilon |B|$, we have
$|\dens(A',B')-\dens(A,B)|\leq \epsilon$.  A partition
$V_1,\ldots,V_k$ of a set into disjoint parts is called an
\emph{equipartition} if $||V_i|-|V_j||\leq 1$ for $1\leq i<j\leq k$.

Szemer\'edi's Regularity Lemma~\cite{sz} states that for every real
$\epsilon>0$ and integer~$m\geq 1$ there exist two integers $M$ and
$n_0$ with the following property.  For every graph $G$ with
$n\geq n_0$ vertices there exists an equipartition of the vertex set
into $k$ parts $V_1,\ldots, V_k$, $m\leq k\leq M$, such that all but
at most~$\epsilon k^2$ of the pairs~$(V_i,V_j)$ are
$\epsilon$-regular.

Irregular pairs cannot be avoided as witnessed by the example of
half-graphs, see~\cite{alon1994algorithmic}. A \emph{half-graph} (or
\emph{ladder}) of order $k$ is the bipartite graph $H_k$ on vertices
$a_1,\ldots, a_k$ and $b_1,\ldots, b_k$ with edges $\{a_i, b_j\}$ for
$i\leq j$.  For disjoint subsets $X,Y$ of vertices of a graph $G$, we
write $G[X,Y]$ for the subgraph of $G$ semi-induced by $X$ and $Y$,
that is, the subgraph with vertex set $X\cup Y$ and all the edges of
$G$ with one endpoint in $X$ and one endpoint in $Y$. A bipartite
graph $H$ is a semi-induced subgraph of $G$ if $H$ is isomorphic to
$G[X,Y]$ for some disjoint subsets $X$ and $Y$ of $V(G)$.

Malliaris and Shelah~\cite{regularity_lemmas} proved that semi-induced
half-graphs are the only reason for the exceptional pairs and proved a
stronger regularity lemma (see \cite{jiang2020regular} for an in-depth
discussion of regularity lemmas for restricted graph classes). By
giving up the requirement of partitioning into a constant number of
parts, they proved that for every class $\CC$ of graphs that exclude a
half-graph of order $k$ as semi-induced subgraph there exist an
integer $n_0$ and a real $\delta$, such that for every graph $G\in\CC$
of size $n \geq n_0$ there exists an equipartition into disjoint parts
of size $\Omega(n^\delta)$ such that after possibly omitting one
vertex from each part the following holds.

\vspace{-2mm}
  \begin{enumerate}[label=(\arabic*)]
  \item Every part forms a clique or an independent set.
  \item For every two parts $A$ and $B$, either for all vertices
    $a\in A$, except for at most $2k$, it holds that
    $|N(a)\cap B| \leq 2k$, or for all $a\in A$, except for at most
    $2k$, it holds that $|N(a)\cap B| \geq |B|-2k$.
  \end{enumerate}
  \vspace{-1mm}

\begin{figure}
  \begin{center}
    \includegraphics[scale=0.75]{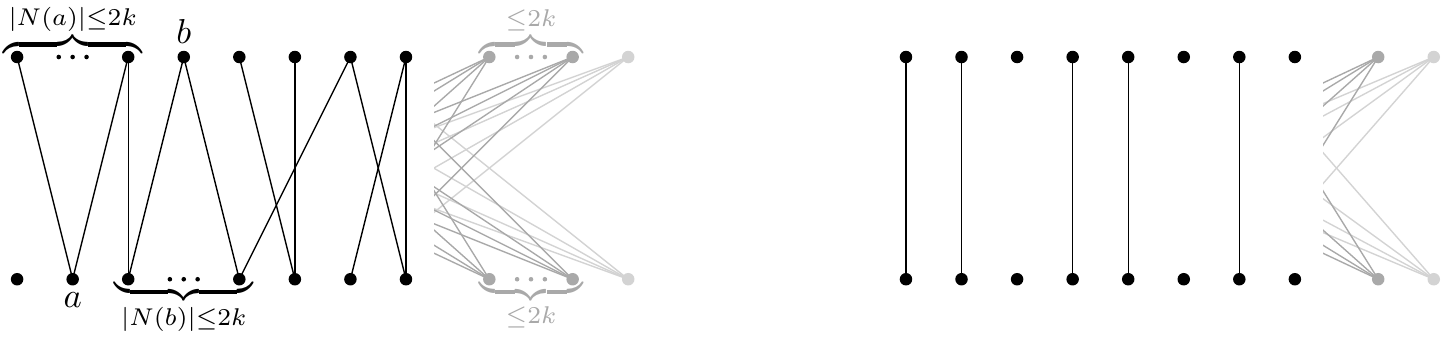}
  \end{center}
  \caption{A possible connection between two parts for Malliaris and Shelah's regularity lemma (left) and our regularity lemma for monadically stable classes (right).
      In this example in both cases, two parts $A$ and~$B$ form an independent set and are placed at the top and bottom,
      the local exceptional vertices ($2k$ or one) of each part are placed on the right in dark gray,
      and the additional single global exceptional vertex of each part is placed even further right in light gray.
      The defining statements are applied once from $A$ to $B$ and once from $B$ to $A$.
    }
  \label{fig:regularity-comparison}
\end{figure}

  For every monadically stable class of graphs $\CC$ we strengthen
  this result by replacing the at most~$2k$ exceptional vertices by
  single exceptional vertices. We prove that there exist an integer
  $n_0$ and a real~$\delta$, such that for every graph $G\in\CC$ of
  size $n \geq n_0$ there exists an equipartition into disjoint parts
  of size~$\Omega(n^\delta)$ such that after possibly omitting one
  vertex from each part the following holds (see \Cref{fig:regularity-comparison} for an illustration of both
  regularity lemmas).
  \begin{enumerate}[label=(\arabic*)]
  \item Every part forms a clique or an independent set.
  \item For every two parts $A$ and $B$, either for all vertices
    $a\in A$, except for at most one, it holds that
    $|N(a)\cap B| \leq 1$, or for all $a\in A$, except for at most
    one, it holds that $|N(a)\cap B| \geq |B|-1$.  This means, with
    the exception of one vertex per part, the connection between $A$
    and $B$ semi-induces either a subgraph of a matching or a
    supergraph of a co-matching.
  \end{enumerate}
  
  \setlength{\belowcaptionskip}{-10pt}

\smallskip
\noindent
\textbf{Polynomial kernels for independent and dominating
set in powers of nowhere dense graphs.}
Finally, we study algorithmic applications of structurally nowhere dense classes.
A parameterized graph problem is called fixed-parameter tractable with
respect to a parameter~$k$ if on input $G$ and~$k$ it can be solved in
time $f(k)\cdot |V(G)|^c$ for some function $f$ and constant~$c$.  The
independent set and dominating set problem are important algorithmic
problems, which are unlikely to be fixed-parameter tractable in
general graphs~\cite{downey1995fixed}. A kernel is a polynomial time
algorithm that reduces an instance $(G,k)$ to an equivalent instance
$(H,k')$ such that $|H|+|k'|$ is bounded by a function $g(k)$ only.
In case $g$ is a polynomial function, one speaks of a polynomial
kernel.  For an integer $d$ and a graph $G$, the $d$th power of $G$ is
the graph $G^d$ on the same vertex set as $G$, where two (distinct)
vertices are connected if their distance in $G$ is at most~$d$. A
class $\Dd$ is a power of another class $\Cc$ if there exists $d$ such
that $\Dd=\{G^d~:~G\in \Cc\}$. It was recently shown that the
independent set problem and dominating set problem are fixed-parameter
tractable on powers of nowhere dense graphs~\cite{fabianski2019},
however no polynomial kernels were known.  We complete the picture by
proving that both problems admit polynomial kernels on powers of
nowhere dense classes of graphs.

Note that the independent/dominating set problem on the $d$th power
$G^d$ of a graph $G$ naturally corresponds to the following
distance-$d$ version of the problem on $G$. A \emph{distance-$d$}
independent set in a graph $G$ is a set of vertices of pairwise
distance greater than~$d$, while a \emph{distance-$d$ dominating} set
is a set of vertices such that every vertex of $G$ has distance at
most $d$ to a vertex of the dominating set.
The distance-$d$ independent set problem and distance-$d$ dominating
set problem are fixed-parameter tractable on every nowhere dense class
of graphs~\cite{DawarK09}, and in fact admit polynomial
kernels~\cite{EickmeyerGKKPRS17,kreutzer2018polynomial}. However, note
that our result is not implied by these results, as the difficulty in
dealing with powers of nowhere dense classes stems from the fact that
it is hard to compute for a given graph $H$ a graph $G$ such that
$H=G^d$.

\section{Preliminaries}\label{sec:prelims}

We use standard notation from graph theory and model theory
and refer to~\cite{Diestel} and~\cite{Hodges} for extensive background. 
We write $[m]$ for the set of integers $\{1,\ldots,m\}$. 

\smallskip
\noindent
\textbf{Graphs.}
Given a graph $G$, we denote its vertex set by $V(G)$ and its edge set by $E(G)$. 
For a vertex $u \in V(G)$ we define $N(u) := \{v~|~\{u,v\} \in E(G)\}$ to be the \emph{open neighborhood} of $u$.
A \emph{partition} of~$V(G)$ is a collection of pairwise disjoint, non-empty subsets whose union is $V(G)$. 
An \emph{$r$-subdivision} of $G$ is a graph obtained by subdividing each edge of $G$ exactly $r$ 
times. 
We call a bipartite graph $H$ a \emph{semi-induced subgraph} of $G$ if $H$ can be obtained from $G$ by choosing two disjoint subsets $A,B$ of $V(G)$, taking the subgraph of $G$ induced by $A\cup B$ and removing all edges with both endpoints in $A$, respectively $B$.
Let $H$ be a bipartite graph with vertices 
$a_1,\ldots, a_n$ and $b_1,\ldots, b_n$. Then $H$ is
\begin{itemize}
\item a \emph{biclique} of order $n$ if we have $\{a_i, b_j\}\in E(H)$ for all $i,j\in [n]$,
\item a \emph{matching} of order $n$ if we have $\{a_i, b_j\}\in E(H)\Longleftrightarrow
i=j$ for all $i,j\in [n]$, 
\item a \emph{co-matching} of order $n$ if we have $\{a_i,b_j\}\in E(H)\Longleftrightarrow i\neq j$ for all $i,j\in [n]$, and
\item a \emph{half-graph} or \emph{ladder} of order $n$ if $\{a_i,b_j\}\in E(H) \Longleftrightarrow i\leq j$ for all $i,j\in [n]$. 
\end{itemize}

\smallskip
\noindent
\textbf{Monadic stability.}
A \emph{transduction} (on graphs) is an operation mapping an input graph to a set of output graphs.
A transduction is defined by an integer $k$ and two first-order formulas $\phi(x,y),\nu(x)$ over the language of vertex colored graphs, where $\phi$ is a symmetric and irreflexive formula (that is $\models \phi(x,y)\leftrightarrow \phi(y,x)$ and $\models \neg\phi(x,x)$) and is the composition of the following operations.
First, the input graph is mapped to $k$ disjoint copies.
Edges between the copies are created by pairwise connecting copies of the same vertex.
The resulting graph is mapped to the set of all of its possible vertex colorings.
At last, every colored graph $G^+$ is mapped to an uncolored graph $H$ whose vertex set $\{v\in V(G^+) ~|~ G^+\models \nu(v) \}$ is defined by $\nu$ and
whose edge set
$\{\{u,v\} ~|~ G^+\models \phi(u,v), u,v \in V(H)\}$ is defined by $\phi$. Note that since~$\phi$ is symmetric and irreflexive, it defines the edge set of an undirected graph. 

This definition lifts to classes of graphs where we say a class $\CC$ \emph{transduces} a class $\DD$ if there exists a fixed transduction $\mathsf T$ which produces from $\CC$ all graphs from $\DD$, i.e. $\DD \subseteq \bigcup_{G \in  \CC} \mathsf T(G)$.
We call a class \emph{monadically stable} if it does not transduce the class of all ladders.
Generalizing this notion, we call a class \emph{monadically NIP} if it does not transduce the class of all graphs.

Transductions are transitive, i.e., for classes $\CC_1,\CC_2,\CC_3$ we have that if $\CC_1$ transduces $\CC_2$
and $\CC_2$ transduces $\CC_3$, then also $\CC_1$ transduces $\CC_3$.
Combining this observation with the fact that there exists a simple transduction yielding all induced subgraphs of an input graph, 
we will from now on assume without loss of generality that every monadically stable class is \emph{hereditary}, i.e., closed under taking induced subgraphs.

  


\smallskip
\noindent
\textbf{Indiscernible sequences.}
Let $G$ be a graph and $\phi(x_1,\ldots, x_m)$ a formula. 
A sequence $(v_i)_{1\leq i\leq n}$ of vertices from $G$ 
is a \mbox{\emph{$\phi$-indis\-cernible sequence}} of 
length $n$, if for every two sequences of indices
  $i_1 < \dots < i_m$ and
  $j_1 < \dots < j_m$ from $[n]$ we have 
  $
  G \models 
  \phi(v_{i_1}, \ldots, v_{i_m}) 
  $
  if and only if
  $
  G \models \phi(v_{j_1}, \ldots, v_{j_m}).
  $
For a set of formulas $\Delta$ we call a sequence
\emph{$\Delta$-indiscernible} if it is $\phi$-indiscernible for 
all $\phi\in \Delta$. 
In classical model theory, indiscernible sequences are usually defined over arbitrary structures, where they have infinite length and may consist of tuples of elements.
However for our purpose we can restrict ourselves to finite sequences of vertices in graphs.

Indiscernible sequences have recently gained attention also in finite model theory and graph theory when Malliaris and Shelah showed that in \emph{stable} structures we can find polynomially large indiscernible sequences~\cite[Theorem 3.5]{regularity_lemmas}.
The extraction process was made algorithmic in \cite[Theorem 2.8]{kreutzer2018polynomial} and we now use the formulation of \cite{dreier2022iaw}.

\begin{theorem}\label{thm:extraction}
  Let $\CC$ be a stable class of graphs and let $\Delta$ be a finite set of first-order formulas.
  Assume that we can evaluate, given $\phi \in \Delta$, an $n$-vertex graph 
  $G \in \CC$  and a tuple~$\bar v$ in $G$, whether $G \models \phi(\bar v)$ in time  
  $f(\CC,\Delta,n)$.
  Then there is an integer $t$ and an algorithm that 
  finds in any vertex-sequence of size at least $m^t$ in an $n$-vertex graph $G \in\CC$ 
  a $\Delta$-indiscernible sub-sequence of length $m$.
  The running time of the algorithm is 
  $\Oof(|\Delta| \cdot k \cdot n^{1/k} \cdot f(\CC,\Delta,n))$, where $k$
  is the maximal number of free variables of formulas in~$\Delta$.
\end{theorem}

We refrain from formally defining the notion of stability and 
just note that every monadically stable class is also stable, hence, the theorem is applicable in our context. 

\medskip\noindent
For $k\in\NN$ we define 
$
\Delta_k := 
\{E(x,y)\}
\cup
\{
\gamma^S_k
~|~
S\subseteq[k]\}
$
where for every
$S\subseteq[k]$ we define
\[
\gamma^S_k(y_1,\ldots,y_k) 
:=
\exists z \big(
\bigwedge_{i \in S} E(z,y_i) 
\bigwedge_{i \notin S} \neg E(z,y_i)\big).
\]
Since for every graph $G$, set $S\subseteq [k]$ and tuple $\bar v \in V(G)^k$, we can evaluate $G\models \gamma^S_k(\bar v)$ in time $\Oof(k\cdot n)$, we derive the following corollary of \cref{thm:extraction}.

\begin{corollary}\label{cor:delta_extraction}
For every stable class of graphs $\CC$ and integer $k$ there exists an integer $t$ and an algorithm that, given an $n$-vertex graph $G\in \CC$ and a 
sequence $J \subseteq V(G)$ of length $m^t$, extracts a $\Delta_k$-indiscernible sub-sequence of $J$ of length $m$ in time $\Oof(2^k \cdot k^2 \cdot n^2)$. 
\end{corollary}

Note that since the formula $E(x,y)$ is included in $\Delta_k$, every $\Delta_k$-indiscernible sequence is either a clique or an independent set.
In \cite{dreier2022iaw}, 
Dreier, M\"ahlmann, Siebertz, and Toru\'nczyk 
studied the behavior of indiscernible sequences in monadically stable and monadically NIP classes of arbitrary structures.
Their results are of very general nature and work for arbitrary formulas.
For our purposes we restrict their results to monadically stable classes of graphs using only the edge relation and distance formulas.
Here we have that $\Delta_k$-indiscernible sequences impose a strong structure on their neighborhood.
The following follows from \cite[Corollary 3.5]{dreier2022iaw}.

\begin{theorem}\label{thm:neighborhood}
  For every monadically stable class of graphs $\CC$ there exist integers $n,k$ such that for every $\Delta_k$-indiscernible sequence $I$ of length at least $n$ in a graph $G\in\CC$ and for every vertex $v\in V(G)$ we have that
  $
    |N(v) \cap I| \leq 1 \text{ or } |N(v) \cap I| \geq |I|-1.
  $
\end{theorem}

They also prove strong properties for sequences of disjoint first-order definable sets.
In our case we are only interested in disjoint balls of radius $r$ around vertices.
The following follows from \cite[Theorem 4.2 and proof of Theorem 5.2]{dreier2022iaw}.

\begin{theorem}\label{thm:families}
  For every monadically stable class of graphs $\Cc$ and integer $r$, there exist integers $k$ and~$t$ such that 
  for every $m\in \NN$
  every graph $G \in \CC$ and every set $J$ of size $m^t$ of disjoint $r$-balls in $G$ the following holds.
  There exists a subset $I \subseteq J$ of size $m$ with the following two properties.

\vspace{-2mm}
\begin{enumerate}
\item There exists a set $S$ of at most $k$ sample vertices such that for every $v\in V(G)$ there exists a sample vertex $s(v) \in S$ and an exceptional ball $X(v)\in I$ such that 
    for all
    $B \in I \setminus \{X(v)\}$ 
    it holds that 
    $N(v) \cap B = N(s(v)) \cap B$.

\item There exists a coloring $\mathrm{col}: V(G) \rightarrow [k]$ with $k$ colors and a relation $R \subseteq [k]^2$ such that for every two vertices $u,v$ from different $r$-balls in $I$ we have that
    $G\models E(u,v)$
    if and only if
    $\big(\mathrm{col}(u),\mathrm{col}(v)\big) \in R$.
\end{enumerate}
\end{theorem}
\vspace{-2mm}
In the above theorem we can assume that for every sample vertex $s\in S$ we have that $N(s)$ intersect with every or none of the $r$-balls from $I$.
This can be achieved by adding the vertices of $S$ as constants to $G$ and taking a $\Psi$-indiscernible subsequence $I'$ of $I$ where $\Psi(y) := \{\psi_s(y) = \exists z \in N(s): \mathrm{dist}_{\leq r}(y,z)~|~s\in S\}$.
By \cref{thm:extraction} the size of $I'$ will still be polynomial.


\section{Ramsey numbers in monadically stable graph classes}

The Ramsey number
$R(s,t)$ denotes the smallest number of vertices that a graph must have such
that one can either find a clique of size at least $s$ or an independent set of size at least $t$.
As shown by Ajtai, Koml{\'o}s and Szemer{\'e}di~\cite{ajtai1980note} we
have $R(s,t)\in \mathcal{O}_s(t^{s-1}/(\log t)^{s-2})$. We improve these
bounds in monadically stable classes by showing that for every fixed $s \ge 3$ and $K_s$-free monadically
stable class $\Cc$ of graphs, there exists a constant $\delta>0$ such that
in $\Cc$ we have $R(s,t)\in \mathcal{O}_{\Cc}(t^{s-1-\delta})$.
The main result of this section is the following theorem.

\begin{theorem}\label{thm-ramsey}
  Let $s \ge 3$ and $\CC$ be a $K_s$-free monadically stable class of graphs.
  There exists a real $\delta > 0$ and an integer $n_0$ such that every graph $G\in\CC$ with $n\geq n_0$ vertices contains an independent set of size at least $n^{\frac{1}{s-1}+\delta}$.
\end{theorem}

The aforementioned bound $R(s,t)\in \mathcal{O}_{\Cc}(t^{s-1-\delta})$ for monadically stable classes
then follows from the following remark.

\begin{remark}\label{remark:Rst}
Assume we have a graph class $\CC$, integers $s \ge 3$ and $n_0$ and a real $0 < \delta \le 1$ such that
every $K_s$-free graph from $\CC$ with $n\geq n_0$ vertices contains an independent set of size at least $n^{\frac{1}{s-1}+\delta}$, that is, $R(s,n^{\frac{1}{s-1}+\delta}) \le n$ for all $n \ge n_0$ is true in $\Cc$.
By setting $t=n^{\frac{1}{s-1}+\delta}$, we obtain 
$n=t^\frac{s-1}{1+(s-1)\delta}$.
Basic arithmetic
then yields $\frac{s-1}{1+(s-1)\delta} \le s-1-\delta$ for $0 < \delta \le 1$, $s \ge 3$
and therefore $n \le t^{s-1-\delta}$. This means on~$\CC$ we have $R(s,t) \le t^{s-1-\delta}$.
\end{remark}

The proof of \Cref{thm-ramsey} will be by induction on $s$. The heart of the proof is the base case $s=3$, where $\Cc$ is triangle-free.

\begin{lemma}\label{lem:triangleFree}
For every triangle-free monadically stable class $\CC$ of graphs there exists a real $\delta > 0$ and an integer $n_0$ such that every graph $G\in\CC$ with $n\geq n_0$ vertices contains an independent set of size at least $n^{\frac{1}{2}+\delta}$.
\end{lemma}

The core of the proof of \cref{lem:triangleFree} is to compute indiscernible sequences and reason about their neighborhoods using \cref{thm:neighborhood} and \cref{thm:families}.
For technical reasons we want the additional property that the indiscernible sequences have a ``large'' neighborhood. We first show that if we cannot find such indiscernible sequences, then the graph contains a large independent set.

\begin{proof}[Proof of \Cref{lem:triangleFree}]
    Let us fix a graph $G \in \CC$. We greedily build an independent set $\cal I$ as follows.
    We look for an independent set $I$ of size $\lceil (n/2)^\delta \rceil$ with $|N(I)\setminus I| \le n^{\frac{1+\delta}{2}}$,
    where the value $0 < \delta \le 1$ depends only on $\CC$ and will be specified later.
    Then we place the vertices of $I$ into the independent set $\cal I$ and remove $I$ and its neighborhood from the graph, since these vertices are not allowed to be chosen into~$\cal I$ anymore.
    We repeat this process as long as possible.
    Let us first show that the statement of the lemma follows if we
    can repeat this process as long as there are still at least $n/2$ vertices in the graph.
    In each iteration we remove at most $\lceil (n/2)^\delta \rceil + n^{\frac{1+\delta}{2}}$ vertices
    from the graph, add $\lceil (n/2)^\delta \rceil$ vertices to the independent set $\cal I$ and do not terminate unless $n/2$ vertices are removed.
    We choose $n_0$ to be large enough such that with $n \ge n_0$ it holds that
    $4\lceil (n/2)^\delta \rceil + 4 n^{\frac{1+\delta}{2}} \le n^\frac{1+1.1\delta}{2}$.
    The process therefore yields an independent set $\cal I$ of size at least
    $$
    \lceil (n/2)^\delta \rceil \cdot \frac{n/2}{\lceil (n/2)^\delta \rceil + n^{\frac{1+\delta}{2}}} \ge
    \frac{n^{1+\delta}}{4 \lceil (n/2)^\delta \rceil + 4 n^{\frac{1+\delta}{2}}} \ge
    \frac{n^{1+\delta}}{n^\frac{1+1.1\delta}{2}} =
    n^{\frac{1+0.9\delta}{2}}
    $$
    and the statement of the lemma follows.

    From now on, we can therefore assume that the process could not proceed, while there were still at least $n/2$ vertices in the graph.
    Thus, we have a subgraph $G'$ of $G$ with at least $n/2$ vertices and every independent set $I$ in $G'$
    of size $\lceil (n/2)^\delta \rceil$ has $|N(I)\setminus I| \ge n^{\frac{1+\delta}{2}}$ (later on referred to as assumption $(*)$).

    We search for a $\Delta_{k_1}$-indiscernible sequence $I_0$ in $G'$ with $k_1$ chosen as required by \cref{thm:neighborhood}.
    By \cref{cor:delta_extraction}, $I_0$ can be chosen of length at least $(n/2)^{\delta_1}$ for some real $\delta_1$ depending only on $\CC$.

    By triangle-freeness, $I_0$ is not a clique but an independent set.
    Let $G'_1$ be the induced subgraph of $G'$ where all vertices adjacent to at least two elements of $I_0$ have been removed.
    The radius $1$-balls around the elements of $I_0$ in $G'_1$ are disjoint and we can apply \cref{thm:families}, yielding a
    real $\delta_2$ and an integer $k_2$, both depending only on $\CC$, with the following properties.
    There exists a subset $I \subseteq I_0$ of size at least $(n/2)^{\delta}$ for $\delta := \delta_1 \cdot \delta_2$ together with a $k_2$-coloring $\KK$ of the $1$-balls around $I$ in $G'_1$, completely determining the edge relation between vertices from different balls.

    We go back to the graph $G'$ where we partition $N(I)$ into two sets $B_1 \cup B_2$, where $B_1$ contains all vertices that are adjacent to exactly one vertex in $I$ and
    $B_2$ contains all vertices that are adjacent to at least two vertices in $I$.
    Note that in $G'$ the coloring $\KK$ still describes the adjacency between elements from $B_1$ with different neighbors in $I$.
    Since $I$ is a sub-sequence of the $\Delta_{k_1}$-indiscernible sequence $I_0$,  \Cref{thm:neighborhood} still applies and all vertices in $B_2$ are adjacent to at least $|I|-1$ vertices of $I$.
    This means any two vertices $u,v \in B_2$ have a common neighbor in $I$
    and since $G'$ is triangle-free, there can be no edge between $u$ and $v$,
    and $B_2$ forms an independent set.
    We can assume $|B_2| \le \frac{1}{2} n^{\frac{1+\delta}{2}}$, since otherwise the statement of the lemma follows.
    As argued before $I$ is also an independent set.
    Thus, by assumption $(*)$, $|N(I)\setminus I| \ge n^{\frac{1+\delta}{2}}$.
    In summary, we so far have $N(I)\setminus I = B_1 \cup B_2$, $|N(I) \setminus I| \ge n^{\frac{1+\delta}{2}}$ and $|B_2| \le \frac{1}{2}n^{\frac{1+\delta}{2}}$.
    This means $|B_1| \ge \frac{1}{2} n^{\frac{1+\delta}{2}}$.

    We pick the color class $K \in \KK$ containing the most vertices of $B_1$.
    It has size at least $|K| \ge |B_1|/k_2 \ge \frac{1}{2k_2} n^{\frac{1+\delta}{2}}$.
    For $v \in I$ let $K_v = K \cap N(v)$.
    Since every vertex in $B_1$ has exactly one neighbor in $I$, this describes a partition of $K$.
    For every $v \in I$, the vertices in $K_v$ have a common neighbor and thus form (by triangle-freeness of $G$) an independent set.
    If the relation $R$ does not contain $(K,K)$ then vertices from different sets $K_v$ and $K_w$ are not connected for all $v \neq w \in I$
    and therefore $K$ forms an independent set.
    On the other hand,
    if the relation $R$ contains $(K,K)$ then vertices from different sets $K_v$ and $K_w$ are connected for all $v \neq w \in I$.
    Then there can be at most two vertices $v \in I$ with $K_v \neq \varnothing$, since otherwise there would be a triangle.
    Hence, there exists $v \in I$ with $|K_v| \ge |K|/2$.
    In either case, we find an independent set of size at least
    $|K|/2 \ge \frac{1}{4k_2} n^{\frac{1+\delta}{2}}$.
    The statement of the lemma now follows.
\end{proof}

We now turn to the proof of \cref{thm-ramsey}. The proof is by induction with \cref{lem:triangleFree} as the base case.

\begin{proof}[Proof of \Cref{thm-ramsey}]
    If $s=3$, the claim follows from \Cref{lem:triangleFree}.
    For larger values of $s$, we prove the claim by induction.
    The class $\CC'$ of $K_{s-1}$-free induced subgraphs of $\CC$ is monadically stable.
    By induction, there exists $0.1 \ge \delta' > 0$ and $n'_0$
    such that every graph in $\CC'$ with $n\geq n'_0$ vertices contains an independent set of size at least $n^{\frac{1}{s-2}+\delta'} \ge n^{\frac{1+\delta'}{s-2}}$.
    Let $G \in \CC$ be a graph of size $n \ge n_0 \ge n'_0$, where we specify $n_0$ later.
    Let $\beta=(1-\delta'/2)\frac{s-2}{s-1}$.
    We distinguish two cases.

    First, assume there is a vertex $v$ with degree at least $n^\beta$.
    The induced subgraph $G[N(v) \setminus \{v\}]$ has size at least $n^\beta$ and is $K_{s-1}$-free.
    Since it is contained in $\CC'$,
    there is an independent set in $G[N(v) \setminus \{v\}]$ (and therefore also in $G$) of size at least $n^{\beta \frac{1+\delta'}{s-2}}$.\\
    Since $\delta' \le 0.1$, we have
     $(1-\delta'/2)(1+\delta') = 1 + \delta'/2 - \delta'^2/2 \ge 1 + \delta'/3$
    and therefore:\\
    $$
    \beta \frac{1+\delta'}{s-2}
    =
    (1-\delta'/2) \cdot \frac{s-2}{s-1} \cdot \frac{1+\delta'}{s-2}
    =
    (1-\delta'/2)(1+\delta') \cdot \frac{1}{s-1} \ge \frac{1}{s-1} + \frac{\delta'/3}{s-1}.
    $$
    We set $\delta = (\delta'/3)/(s-1)$.
    Then $G$ contains an independent set of size at least $n^{\frac{1}{s-1}+\delta}$.

    If on the other hand there is no vertex with degree at least $n^\beta$ then all vertices have degree less than~$n^\beta$.
    In this case, we greedily construct an independent set by repeatedly picking a vertex and removing it together with its neighbors from the graph.
    This yields an independent set of size at least $n/(n^\beta+1)$.
    We choose $n_0$ to be large enough such that with $n \ge n_0$ it holds that
    $n^\beta+1 \le n^{(1-\delta'/3)\frac{s-2}{s-1}}$.
    Then $G$ contains an independent set of size at least
    $n/(n^\beta+1) \ge n^{1 - (1-\delta'/3)\frac{s-2}{s-1}} = n^{\frac{1}{s-1} + (\delta'/3) \frac{s-2}{s-1}}$.
    The claim follows with~$\delta=(\delta'/3) \frac{s-2}{s-1}$.
\end{proof}


\section{Subdivided bicliques}

For every $r,s,t\in\NN$ denote by $\sdc{r}{s,t}$ the $r$-subdivision 
of the complete bipartite with partitions of size~$s$ and $t$.
In this section we prove the following theorem. 

\begin{theorem}
  For every monadically stable class $\CC$ and every integer $r$,
  there exists an integer $n_0$ and a real $\delta > 0$, such that for every $n\geq n_0$ it holds that every graph
  $G \in \CC$ that contains
  $\mathrm{sd}_{r}(K_{n,n})$ as a subgraph,
  also contains $K_{\lceil n^\delta \rceil , \lceil n^\delta \rceil}$ as a subgraph.
\end{theorem}

\newcommand{\en}{t}

\begin{proof}
  Applying \cref{thm:families} to the class $\CC$ and radius $r$ yields a sample set size bound $k \in \NN$ and 
  an exponent $\frac{1}{\lambda}\in \NN$ bounding the size of the required sequence.
  Define $\en_0 := n$ and $\en_{i+1} := \frac{1}{2} \en_i^\lambda$.
  Choose $n_0$ and $\delta$ such that it holds for all~$n\geq n_0$ and $0\leq i \leq r + 1$ that $k\cdot n^\delta \leq \frac{1}{2} \en_i$.

  We work in the induced subgraph $H$ of $G$ that only contains the vertices of the $\mathrm{sd}_{r}(K_{n,n})$.
  Let $A$ and~$B$ be the sets of principle vertices in the $\mathrm{sd}_{r}(K_{n,n})$.
  Assume towards a contradiction, that~$H$ contains no $K_{n^\delta,n^\delta}$ as a subgraph.

  \begin{claim}
    For every $0\leq i \leq r + 1$, $H$ contains an induced subgraph $H_i$ that contains as a subgraph~$\sdc{r}{\en_i,\en_i}$, with sets of 
	principle vertices $A_i$ and $B_i$ such that the vertices from $A_i$ have disjoint distance-$i$ neighborhoods.
  \end{claim}

  Proving the above claim will immediately lead to a contradiction, as the distance-$(r+1)$ neighborhoods of any two principle vertices in $A_{r+1}$ overlap in 
  every vertex of $B_{r+1}$ as they are the principle vertex of an $r$-subdivided biclique.
  We prove the claim by induction on $i$. 

  \begin{claimproof}

  For the base case set $H_0 := H$, $A_0 = A$, and $B_0 = B$.
  For the inductive step we are given $H_i$
  such that the vertices of $A_i$ have disjoint distance-$i$ neighborhoods.
  They therefore form the centers of a set $J$ of disjoint $i$-balls and we can apply~\cref{thm:families}.
  We obtain a subset $I \subseteq J$ of size $|A_i|^\lambda$ together with a set $S$ of at most $k$ sample vertices such that for every $v \in H_i$, there exists 
  a sample $s(v)\in S$ and an exceptional ball such that $N(v) \cap B = N(s(v))$ for every ball $B \in I$ which is not the exceptional ball.\linebreak 
  Let $s^{-1}(s) := \{v\in V(H_i)~|~s(v) = s\}$ be the set of all vertices described by $s\in S$.
  Let $S_I$ be subset of sample vertices from $S$ whose neighborhood intersects all of the balls from $I$.
  Conversely let $S_\varnothing$ be subset of sample vertices from $S$ whose neighborhood intersects none of the balls from $I$.
  As argued in~\cref{sec:prelims} following~\cref{thm:families} we have that $S = S_I \cup S_\varnothing$. 
  We note that one can also assume, without loss of generality, that $|S_\varnothing| \leq 1$. 

  We now argue that $s^{-1}(s)$ has at most $n^\delta$ elements for every $s\in S_I$.
  Towards a contradiction, assume otherwise and let $Q$ be a subset of $s^{-1}(s)$ of size $n^\delta$.
  Since every vertex from $Q$ can behave exceptionally to at most one ball from $I$ and we have that $|I| \geq 2n^\delta$, we 
  find a subset of balls centered at $I' \subseteq I$ of size~$n^\delta$ in which every vertex from $Q$ has the same neighborhood as $s$.
  Since $s$ has at least one neighbor in every ball of $I'$ this yields a biclique of size $n^\delta$; a contradiction.
  
  It follows that the set $S_I^{-1} := \bigcup_{s\in S_I} s^{-1}(s)$ of vertices described by sample vertices from $S_I$ has size at most $k\cdot n^\delta$.
  Next we build the induced subgraph $H_{i+1}$ of $H_i$ by removing the vertices of $S_I^{-1}$.\linebreak
  Let $A'_i \subseteq A_i$ be the center vertices of the balls in $I$. After the deletion of $S_I^{-1}$ we have that every vertex in $v \in V(H_{i+1})$ is described by a sample vertex from $S_\varnothing$.
  Since the neighborhoods of the vertices in $S_\varnothing$ intersect none of the $i$-balls in $I$, the neighborhood of $v$ can intersect at most one of the $i$-balls in $I$ (i.e. the exceptional ball of $v$).
  It follows that the vertices from $A'_i$ have disjoint $(i+1)$-balls.

  It remains to show that there exist sets $A_{i+1} \subseteq A'_i$ 
  and $B_{i+1} \subseteq B_i$ forming the principle vertices of a 
  $\sdc{r}{\en_{i+1},\en_{i+1}}$ subgraph in $H_{i+1}$.
  We started with the graph $H_i$ containing as a subgraph $\sdc{r}{\en_i,\en_i}$ with principle vertices $A_i$ and $B_i$.
  It follows that $H_i$ contains as a subgraph $\sdc{r}{|A'_i|,\en_i}$ with principle vertices $A'_i$ and $B_i$.
  It holds that for every $t\geq 2$, every graph obtained by removing a single vertex from $\sdc{r}{t,t}$ still contains $\sdc{r}{t-1,t-1}$ as a subgraph.
  Therefore after we removed $S_I^{-1}$ from $H_i$ to build $H_{i+1}$ we still find in it a set $A_{i+1} \subseteq A'_i$ of size
  $
    |A_{i+1}|
    \geq
    |A'_i| - |S_I^{-1}|
    \geq
    |A_i|^\lambda - |S_I^{-1}|
    \geq
    \en_i^\lambda - k \cdot n^\delta
    \geq
    \frac{1}{2} \en_i^\lambda
    =
    \en_{i+1}
  $
  forming one side of a $\sdc{r}{\en_{i+1},\en_{i+1}}$ subgraph in $H_{i+1}$. The other side is formed by a subset of $B_i$.
\end{claimproof}
This completes the proof of the theorem. 
\end{proof}

\section{Regularity in monadically stable classes of graphs}

In this section we prove the following theorem. 

\begin{theorem}\label{thm:regularity-mon-stable}
  For every monadically stable class of graphs $\CC$,
  there exist an integer $n_0$ and a real $\delta$, such that for 
  every graph $G\in\CC$ of size $n \geq n_0$ there exists an equipartition 
  into parts of size $n^\delta$. 
  After possibly omitting one element from each part, they satisfy the following properties:
  \begin{enumerate}[label=(\arabic*)]
    \item Every part forms a clique or an independent set.
    \item For every two parts $A$ and $B$, except for at most one vertex $a_\mathrm{ex} \in A$, 
	for all other vertices $a\in A$, either $|N(a)\cap B| \leq 1$ or $|N(a)\cap B| \geq |B|-1$.  
	This means, with the exception of one vertex per part, the connections between $A$
    and $B$ semi-induce either a subgraph of a matching or a supergraph of a co-matching.
  \end{enumerate}
\end{theorem}

Note that the additional vertex that needs to be omitted from a part $A$ 
when comparing it to a part~$B$ may differ from the vertex that needs to 
be omitted from $A$ when comparing it to a different part $C$.

We follow the approach by Malliaris and Shelah, by partitioning the graph into indiscernible sequences.
We start by analyzing the interaction between two indiscernible sequences.

\begin{lemma} \label{lem:two_indiscernibles}
  For every monadically stable class of graphs $\CC$,
  there exists integers $n_0$ and $k$ such that for every graph $G \in \CC$ the following holds.
  Let $A$ and $B$ be $\Delta_k$-indiscernible sequences of length at least~$n_0$ in $G$. 
  Except for at most one vertex $a_\mathrm{ex} \in A$, for all other vertices $a \in A$, either $|N(a)\cap B| \leq 1$ or $|N(a)\cap B| \geq |B|-1$.
  In particular, after removing at most one vertex from each sequence,
  the connections between $A$ and $B$ semi-induce either a subgraph of a matching or a supergraph of a co-matching.
\end{lemma}

\begin{proof}
Choose $n_0 \geq 5$ and $k$ as required by \cref{thm:neighborhood}.
  
\begin{claim}
The set of vertices of $A$ adjacent to at least two vertices from $B$ has size $0$,$1$, $|A|-1$, or $|A|$.
\end{claim}

\begin{claimproof}
  Assume towards a contradiction that there exist two elements $a_1,a_2 \in A$ adjacent to at least two vertices in $B$ and two vertices 
  $a_3,a_4 \in A$ adjacent to at most one vertex in $B$.
  Since $a_1,a_2$ have at least two neighbors in $B$, by~\cref{thm:neighborhood} they both must have at least $|B|-1$ neighbors in $B$. 
  Therefore there must be a set $B' \subseteq B$ of size at least $|B|-2 \geq 3$ containing only vertices adjacent to both $a_1$ and $a_2$.
  Again by~\cref{thm:neighborhood} every vertex in $B'$ must have at least $|A|-1$ neighbors in $A$.
  As a result either $a_3$ or $a_4$ will have at least two neighbors in $B'$ and therefore in $B$ and we reach a contradiction.  
\end{claimproof}

Let $A' \subseteq A$ be the set of vertices adjacent to at least two vertices from $B$. It has size $0$,$1$, $|A|-1$, or~$|A|$.
Assume $|A'| = 0$.
All vertices of $A$ have at most one neighbor in $B$ and we have at most $|A|$ edges between the two sets. If no two vertices have the same neighbor, then we have a subgraph of a matching between $A$ and $B$ and we are done.
Assume two vertices of $A$ share a neighbor $b$ in $B$.
By \cref{thm:neighborhood}, $b$ must have at least $|A|-1$ neighbors in $A$.
We therefore have at most one edge between $A$ and $B\setminus{\{b\}}$, a subgraph of a matching as required.

If $|A'| = 1$, only one vertex $a \in A$ 
has at least two neighbors in $B$ we omit $a$ from $A$ and argue as in the previous case.
The two cases where $|A'| \geq |A|-1$ follow by symmetry.
\end{proof}

\begin{proof}[Proof of \cref{thm:regularity-mon-stable}]
  Let $n'_0$ and $k$ be from \cref{lem:two_indiscernibles}
  and let $\lambda \leq \frac{1}{2}$ be the real given by \cref{cor:delta_extraction} such that any graph from $\CC$ of size $m$ contains a 
  $\Delta_k$-indiscernible sequence of size at least $m^\lambda$.
  Therefore, we can iteratively build a partition $\GG$ of $\Delta_k$-indiscernible sequences of size $n^{\frac{\lambda}{2}}$ in $G$ as 
  long as $|V(G)\setminus\bigcup\GG| = n - |\bigcup \GG| \geq n^\frac{1}{2}$.
  Once we have that $|\bigcup\GG| > n - n^\frac{1}{2}$,
  we know that $\GG$ contains at least
  \[
    \left \lfloor 
    \frac{n - n^\frac{1}{2}}{n^{\frac{\lambda}{2}}} 
    \right \rfloor
    =
    \left \lfloor 
    n^{1-\frac{\lambda}{2}} - n^{\frac{1}{2}-\frac{\lambda}{2}}
    \right \rfloor
    \geq
    n^\frac{1}{2}
  \] 
  many parts.
  Therefore we can distribute the leftover elements from $V(G)\setminus\bigcup\GG$ among the parts of $\GG$, such that every part of $\GG$ gets at most one additional vertex.
  Now $\GG$ is the desired equipartition with part size at least $n^\delta$ for $\delta := \frac{\lambda}{2}$.
  By omitting the leftover element property $(1)$ follows from the fact that any $\Delta_k$-indiscernible sequence is either a clique or an 
  independent set and property $(2)$ follows from \cref{lem:two_indiscernibles} and by setting $n_0 \geq {n'_0}^{\frac{1}{\delta}}$.
\end{proof}

\section{Polynomial kernels in powers of nowhere dense classes}

In this section we turn to algorithmic applications of indiscernible sequences in monadically stable classes of graphs. 
We focus on powers of classes of nowhere dense graphs.
Recall that the $d$th power $G^d$ of a graph $G$ is obtained by pairwise connecting in $G$ vertices that have distance at most $d$.
This notion is lifted to classes of classes by defining $\CC^d := \{G^d~|~G\in\CC\}$ to be the $d$th power of a class~$\CC$. 

Powers of nowhere dense graphs were studied by Fabiański et al.\ who showed that they admit fixed-parameter tractable algorithms 
for \textsc{Independent Set} and \textsc{Dominating Set} parameterized by solution size $k$~\cite{fabianski2019}.
We will strengthen their result by giving polynomial kernels for both problems and prove the following theorem.

\begin{theorem}
    \label{thm:kernels}
    For every integer $d$ and nowhere dense class $\CC$,
    \textsc{Independent Set} and \textsc{Dominating Set} parameterized by solution size $k$ admit a polynomial kernel on $\CC^d$.
\end{theorem}

To obtain this result we will make use of the following two properties of powers of nowhere dense graphs.
For every integer $d$ and nowhere dense class $\CC$, the class $\CC^d$ (1) is monadically stable, and (2) excludes a fixed co-matching as a semi-induced subgraph.
For every integer $d$ the operation of
taking the $d$th power of a graph is expressible
as a transduction.
Therefore the first property follows trivially from the monadic stability of nowhere dense classes.
The second property was proved by Fabiański et al. in~\cite[Theorem 11]{fabianski2019}.
Assuming those two properties, we will construct our kernels.
In fact, our results apply to all graph classes satisfying these two properties, with powers of nowhere dense classes merely being the most prominent example.

\begin{theorem}\label{thm:kernel_is}
    \textsc{Independent Set} parameterized by solution size $k$ admits a polynomial kernel on every
    monadically stable class of graphs which excludes
    a co-matching of size $\ell$ as a semi-induced subgraph.
\end{theorem}

\begin{proof}
    We search for an independent set of size $k$ in a graph $G$, which we assume to be from a monadically stable class $\CC$ that excludes a co-matching of size $\ell$.
	By~\cref{thm:neighborhood}, for every monadically stable class of graphs $\CC$ there exist integers $n_0$ and $p$ such that for every $\Delta_p$-indiscernible sequence 
	$I$ of length at least~$n_0$ in a graph $G\in\CC$ and for every vertex $v\in V(G)$ we have that $|N(v) \cap I| \leq 1 \text{ or } |N(v) \cap I| \geq |I|-1$. 
	We fix such a pair of integers $n_0$ and $p$, as given by~\cref{thm:neighborhood}. 
	Now by~\cref{cor:delta_extraction}, there exists an integer~$t$ and an 
	algorithm that, given an $n$-vertex graph $G\in \CC$, where $n \geq n_0^t$, extracts a $\Delta_p$-indiscernible sequence $I \subseteq V(G)$ of 
	length~$n_0$ in time $\Oof(2^p \cdot p^2 \cdot n^{2})$. 
    We now have that in any graph $G \in \CC$ of size at least $n_0^t$, we can find in time $\Oof(2^p \cdot p^2 \cdot n^{2})$ a $\Delta_p$-indiscernible sequence $I$ 
	of size at least $n^{1/t}$ such that, by~\Cref{thm:neighborhood}, every vertex of $V(G) \setminus I$ is adjacent to at most one vertex from $I$ 
	or all but at most one vertex from $I$. Recall that since we include the edge relation in $\Delta_p$, every $I$ is either a clique or an independent set. 

    If $n \leq (k+1)^t$, then the graph itself is a polynomial kernel, so we can assume without loss of generality that $n > (k+1)^t$.
    This means that we can find in any graph $G \in \CC$ of size at least~$(k+1)^t$ in time $\Oof(2^p \cdot p^2 \cdot n^{2})$ a $\Delta_p$-indiscernible sequence $I$ 
	of size at least $k + 1$.
    We will (for later) similarly assume, without loss of generality, that $k + 1 \geq \ell$; since otherwise we can solve 
	the problem in polynomial time, i.e., $\Oof(n^{\ell})$-time, and return a trivial yes- or no-instance. 
    Our kernelization algorithm proceeds by applying the following reduction rule exhaustively.
	
    \begin{tcolorbox}[colback=black!3!white,colframe=black!30!white]
      If $G$ has at least $(k + 1)^t$ vertices, find a $\Delta_p$-indiscernible sequence of size at least~$k + 1$.
      If $I$ is an independent set return a trivial yes-instance.
      Otherwise remove from $G$ all vertices of $I$ except for~$k$ many.
    \end{tcolorbox}
	
    Since $\CC$ is closed under taking induced subgraphs, we can apply this rule repeatedly without leaving $\CC$.
    The procedure runs in polynomial time and obviously, upon termination, the resulting graph has size at most $(k + 1)^t$.
    Since the procedure only removes vertices, it cannot turn no-instances into a yes-instances.
    It remains to argue that the procedure does not turn yes-instances into no-instances.

    If $I$ forms an independent set, then we return a trivial yes-instance, so the only remaining case is $I$ forming a clique.
    In this case, $I$ can contain at most one vertex from a solution.
    Assume there exists a vertex outside $I$ that is non-connected to only one element of $I$.
    By indiscernibility of~$I$, there exists such a ``private vertex`` for every element in $I$ which implies a co-matching of size at least~$k + 1 \geq \ell$.
    This is a contradiction to our assumption of excluding a co-matching of size $\ell$.
    Therefore, by \Cref{thm:neighborhood}, every vertex outside $I$ is adjacent to either all or at most one vertex of $I$.

    Assume we have a yes-instance, witnessed by an independent set $S$ of size $k$.
    Since $I$ is a clique, $I$ contains at most one vertex from $S$.
    If $S$ contains no vertex from $I$, then after applying the reduction rule we trivially still have a yes-instance.
    On the other hand, if $S$ contains exactly one vertex from $I$ then the remaining $k-1$ vertices outside $I$ are not adjacent to all vertices in $I$
    and therefore have at most one neighbor each within $I$.
    This means in any subset of $k$ vertices of $I$ there will be at least one vertex that is non-adjacent to the remaining $k-1$
    vertices of $S$, completing it to an independent set of size $k$ and preserving the yes-instance.
\end{proof}

\begin{theorem}\label{thm:kernel_dom}
  \textsc{Dominating Set} parameterized by solution size $k$ admits a polynomial kernel on every
  monadically stable class of graphs which excludes
  a co-matching of size $\ell$ as a semi-induced subgraph.
\end{theorem}

\begin{proof}
  We search for a dominating set of size $k$
  in a graph from a monadically stable graph class $\CC$ which excludes a co-matching of size~$\ell$ as a semi-induced subgraph.
  We outline our approach:
  Given a \textsc{Dominating Set} instance $(G_0,k)$, we will first construct
  an equivalent instance $(G,k)$ of the \textsc{Red Blue Dominating Set} problem.
  The input to this problem is a bipartite graph $G$ whose two parts are colored red and blue, respectively,
  and a number $k$, and the task is to find $k$ blue vertices that dominate all red vertices.
  We will ensure that $G$ is still part of another monadically stable class $\DD$ that excludes a co-matching.
  We then kernelize $(G,k)$ to obtain a kernel $(K,k)$ to \textsc{Red Blue Dominating Set} whose size is polynomial in $k$.
  Finally we will transform $(K,k)$ back into a \textsc{Dominating Set} instance $(K_0,k+1)$ by only introducing $\mathcal{O}(k)$ many new vertices.

  \smallskip\noindent
  \textbf{From \textsc{Dominating Set} to  \textsc{Red Blue Dominating Set}.}
  We create the instance~$(G,k)$ from~$(G_0,k)$ as follows.
  The vertices of $G$ consist of a red copy $A$ and a blue copy $B$ of $V(G_0)$. For every vertex $v \in V(G_0)$ we connect the red copy of $v$ with each blue copy in $N[v]$.
  Note that no vertices of the same color are connected.
  There exists a fixed transduction
  which produces $G$ from $G_0$, so we know that $G$ must be from a monadically stable class $\DD$ depending only on $\CC$.
  It is easy to see that $G_0$ contains a size $k$ dominating set if and only if there exists $k$ vertices from $A$ which together dominate all of $B$.
  Now assume towards a contradiction that $G$ contains a co-matching $C$
  of size $3\ell$ as a semi-induced subgraph.

  We build a semi-induced co-matching $C_0$ in $G_0$ by iteratively taking a red vertex $a$ and a blue vertex $b$ from $C$ that are not connected and adding them to $C_0$.
  Since $a$ and $b$ are non-neighbors in different colors, we know that they are not connected and correspond to different vertices in $G_0$.
  If the blue copy of $a$ or the red copy of $b$ is contained in $C$ we remove them together with their co-matching partners from~$C$.
  In $G_0$, $a$ will be connected to the original of every blue vertex in $C$ except for $b$, 
  and analogously $b$ will be connected to the original of every red vertex in $C$ except $a$.
  In every step, we add one pair to $C_0$ and remove at most three pairs from $C$.
  Therefore by fully exhausting $C$ we find a semi-induced co-matching of size at least $\ell$ in $C$, a contradiction.
  It follows that $G$ contains no co-matching of size $3\ell$.

  \smallskip\noindent
  \textbf{Kernelization.}
  By~\cref{thm:neighborhood}, for every monadically stable class of graphs $\DD$ there exist integers $n_0$ and $p$ such that for every $\Delta_p$-indiscernible sequence 
  $I$ of length at least $n_0$ in a graph $G\in\DD$ and for every vertex $v\in V(G)$ we have that $|N(v) \cap I| \leq 1 \text{ or } |N(v) \cap I| \geq |I|-1$. 
  We fix such a pair of integers~$n_0$ and $p$, as given by~\cref{thm:neighborhood}. 
  Now by~\cref{cor:delta_extraction}, there exists an integer $t$ and an 
  algorithm which given an $n$-vertex graph $G\in \DD$, where $n \geq n_0^t$, extracts a $\Delta_p$-indiscernible sequence $I \subseteq V(G)$ of 
  length~$n_0$ in time $\Oof(2^p \cdot p^2 \cdot n^{2})$. 
  We now have that in any graph $G \in \DD$ of size at least $n_0^t$, we can find in time $\Oof(2^p \cdot p^2 \cdot n^{2})$ a $\Delta_p$-indiscernible sequence $I$ 
  of size at least $n^{1/t}$ such that, by~\Cref{thm:neighborhood}, every vertex of $V(G) \setminus I$ is adjacent to at most one vertex from $I$ 
  or all but at most one vertex from $I$. 
  
  We assume (similarly to~\Cref{thm:kernel_is}), without loss of generality, that $k+2 \ge 3\ell$, $k > 1$, $|B| > (k+2)^t$.
  Thus, if the set of to-be-dominated vertices $B$ contains at least $(k+2)^t$ vertices, there exists a \mbox{$\Delta_p$-indiscernible} sequence $I$ of size at least $k+2$ among $B$
  such that by~\cref{thm:neighborhood}, all vertices in $A$ are connected to at most one or all but at most one vertex in $I$.
  We will kernelize it using the following reduction rules.

  \begin{tcolorbox}[colback=black!3!white,colframe=black!30!white]
    \textbf{1.} If $A$ contains two vertices with the same neighborhood in $B$, remove one of them.\\
    \textbf{2.} If the set of to-be-dominated vertices $B$ contains at least $(k+2)^t$ vertices, find a $\Delta_p$-indiscernible sequence $I$ of size at least $k+2$ among $B$.
      Remove from $G$ all vertices of $I$ except for $k+1$ many.
  \end{tcolorbox}
  The first reduction rule is obviously efficient and safe.
  As discussed above, the second rule can be implemented in polynomial time.
  Let $(G',k)$ be the instance obtained from $(G,k)$ by the second rule, where a $\Delta_p$-indiscernible sequence $I$ is reduced to a sub-sequence $I'$ of size $k+1$.
  If $(G,k)$ is a yes-instance, then removing to-be-dominated vertices from $B$ will still yield a yes-instance.
  To show safety of the second rule, we assume that the reduced instance $(G',k)$ is a yes-instance and show that $(G,k)$ also is a yes-instance.
  Since $(G',k)$ is a yes-instance, there exists a set $S\subseteq A$ of size at most $k$ that dominates all of $I'$.
  As discussed above, by~\cref{thm:neighborhood} every vertex in $S$ is connected to at most one, all, or all but one vertex of $I$.
  Assume for contradiction there exists a vertex connected to all but one vertex of $I$.
  By indiscernibility, one vertex of $I$ having a private non-neighbor would mean that all vertices of $I$ have a private non-neighbor,
  implying the existence of a co-matching of size $|I|\geq k+2 \geq 3\ell$, a contradiction.
  Therefore every vertex in $S$ is either connected to at most one or all vertices of $I$.
  Since $I'$ has $k+1$ vertices and $S$ only $k$, $S$ must include a vertex connected to at least two elements of $I$ and therefore connected to all of $I$.
  Hence, $S$ is also a solution to $(G,k)$, meaning that $(G,k)$ is a yes-instance

  It remains to prove that the kernel $(K,k)$ obtained by exhaustively applying the two rules has size polynomial in $k$.
  Both rules reduce the size of the graph and when applied exhaustively, $B$ has size at most $(k+2)^t$ and $A$ contains no more twins.
  We use the following theorem which applies to \emph{NIP} classes which 
  contain monadically NIP and therefore also monadically stable classes.

  \begin{theorem}[Sauer-Shelah-Lemma \cite{sauer1972sauershelah,shelah1972sauershelah,vapnik1971sauershelah}]\label{thm:sauer_shelah}
    Let $G \in \DD$ and let $B \subseteq V(G)$. 
	Let $\mathcal{P} = \{P_1, P_2, \ldots\}$ be a partition of $V(G) \setminus B$ such that two vertices $u$ and $v$ belong to $P \in \mathcal{P}$ 
	whenever $N(u) \cap B = N(v) \cap B$. If $\DD$ is a NIP class of graphs then there exists a constant $d$ such that $|\mathcal{P}| \leq |B|^{d}+1$. 
  \end{theorem}
  
  Since we have no twins in $A$, every vertex in $a \in A$ is uniquely identified by its neighborhood $N(a) \cap B$
  and therefore $|A| \le |B|^{d}+1 \le (k + 2)^{t \cdot d}+1$.
  Thus, $(K,k)$ has polynomial size.

  \smallskip\noindent
  \textbf{From \textsc{Red Blue Dominating Set} to \textsc{Dominating Set}.}
  Finally we are going to convert $(K,k)$ into a \textsc{Dominating Set} instance $(K_0,k+1)$ by creating a star $K_{1,k+2}$ of size $k+2$ and connecting its center $c$ to every red (i.e. dominating) vertex from $K$.

  If $(K,k)$ admits a red blue dominating set $S$ of size $k$, then $S \cup \{c\}$ is a $k+1$ dominating set for $K_0$.
  Vice versa, the only way to dominate the $k+2$ leaves of $K_{1,k+2}$ with $k+1$ vertices is by picking the center~$c$.
  Hence, if there exists a solution $S$ for $(K_0,k+1)$, it must contain $c$.
  Note that all the red vertices are dominated by $c$ and every blue vertex has no blue neighbor. Therefore, we can replace every blue vertex in $S$
  with any of its red neighbors.
  This yields a set $S'$ containing $c$ and at most $k$ red vertices dominating all blue vertices. These red vertices witness that $(K,k)$ is a yes-instance as well.
\end{proof}

Having proved \cref{thm:kernel_is} and \cref{thm:kernel_dom},
this concludes the proof of \cref{thm:kernels}.



\bibliography{ref}


\end{document}